\documentclass[11pt]{article}
\usepackage{moriond}

\def\Journal#1#2#3#4{{#1} {\bf #2}, #3 (#4)}

\def\PLB{{\em Phys. Lett.}  B}
\def\PRL{\em Phys. Rev. Lett.}
\def\PRD{{\em Phys. Rev.} D}

\def\be{\begin{equation}}
\def\ee{\end{equation}}
\def\bea{\begin{eqnarray}}
\def\eea{\end{eqnarray}}

\newcommand{\Photo}{}

\begin{document}
\vspace*{4cm}
\title{Tevatron Combination and Higgs Boson Properties}

\author{ Weiming Yao\footnote{On behalf of the CDF and D0 Collaborations}} 

\address{Physics Division, Lawrence Berkeley National Laboratory, \\
One Cyclotron Road, Berkeley, CA 94720, USA}

\maketitle\abstracts{
We present the Tevatron combination of searches for the Higgs boson and 
studies of its properties. The searches use up to 10 fb$^{-1}$ of Tevatron 
collider Run II data. We observe a significant excess of events in the mass 
range between 115 and 140 GeV/c$^2$. The local significance corresponds to 
3 Gaussian standard deviations at the mass of 125 GeV/c$^2$. Furthermore, 
we separately combine searches for the Higgs boson decaying to $b\bar b$, 
$\tau^+\tau^-$, $W^+W^-$, and photon pairs in the final states. 
The observed signal strengths in all channels are consistent with the 
presence of a standard model scalar boson with a mass of 125 GeV/c$^2$.
Studies of the couplings at the Tevatron are consistent with SM predictions 
and are complementary to those performed at LHC.}

\section{Introduction}

 The standard model (SM) Higgs boson was first proposed in 1964 as the 
 remnant of a scalar field (H) that could be responsible for the electroweak 
 symmetry breaking~\cite{higgs}. The quest for the Higgs boson particle, the 
 last missing piece of the standard model, has been a major goal of particle 
 physics and a central part of the Fermilab Tevatron collider physics program
 for several decades. Both CDF and D0 collaborations have finished their direct 
 searches for the standard model Higgs boson using the complete Tevatron
 Run II dataset with an integrated luminosity of 10 fb$^{-1}$ of 
 proton-antiproton data. The results reported here are close to the final 
 word on the Higgs studies at the Tevatron, which have been recently 
 submitted for 
 publication~\cite{tevhiggs}. The searches are performed for assumed Higgs 
 masses between 90 and 200 GeV/c$^2$.     

 In the standard model, the mass of the Higgs boson ($m_H$) is a free 
 parameter and is indirectly constrained to be less than 152 GeV/c$^2$ at the 
 95\% conference level (CL)~\cite{ewkfit} by the global fit of the electroweak
 precision data including the recent top-quark mass and $W$ boson mass 
 measurements from the Tevatron~\cite{mtop,mw}. The direct searches 
 from LEP2~\cite{lep} set the limit of Higgs mass 
 above 114.4 GeV/c$^2$ at the 95\% CL. On fourth of July 2012, the 
 ATLAS and CMS collaborations announced the observation of the 
 Higgs-like particle~\cite{atlas,cms} with a mass of 
 approximately 125 GeV/c$^2$. Much of the power of the LHC 
 searches comes from $gg\rightarrow H$ production and Higgs boson decays to 
 $\gamma\gamma$, $W^+W^-$, and $ZZ$, which probes the couplings of the Higgs 
 boson to other bosons. The Tevatron has recently reported a 3.1 sigma 
 excess of $H\rightarrow b\bar b$ events in association with a 
 vector boson production, which may provide the first evidence of Higgs 
 coupling to b quarks~\cite{tevbb}. 

\subsection{Higgs Search Strategies}\label{subsec:prod}

\begin{figure}
\centerline{
\begin{minipage}{0.48\linewidth}
\centerline{
\includegraphics[width=0.95\linewidth]{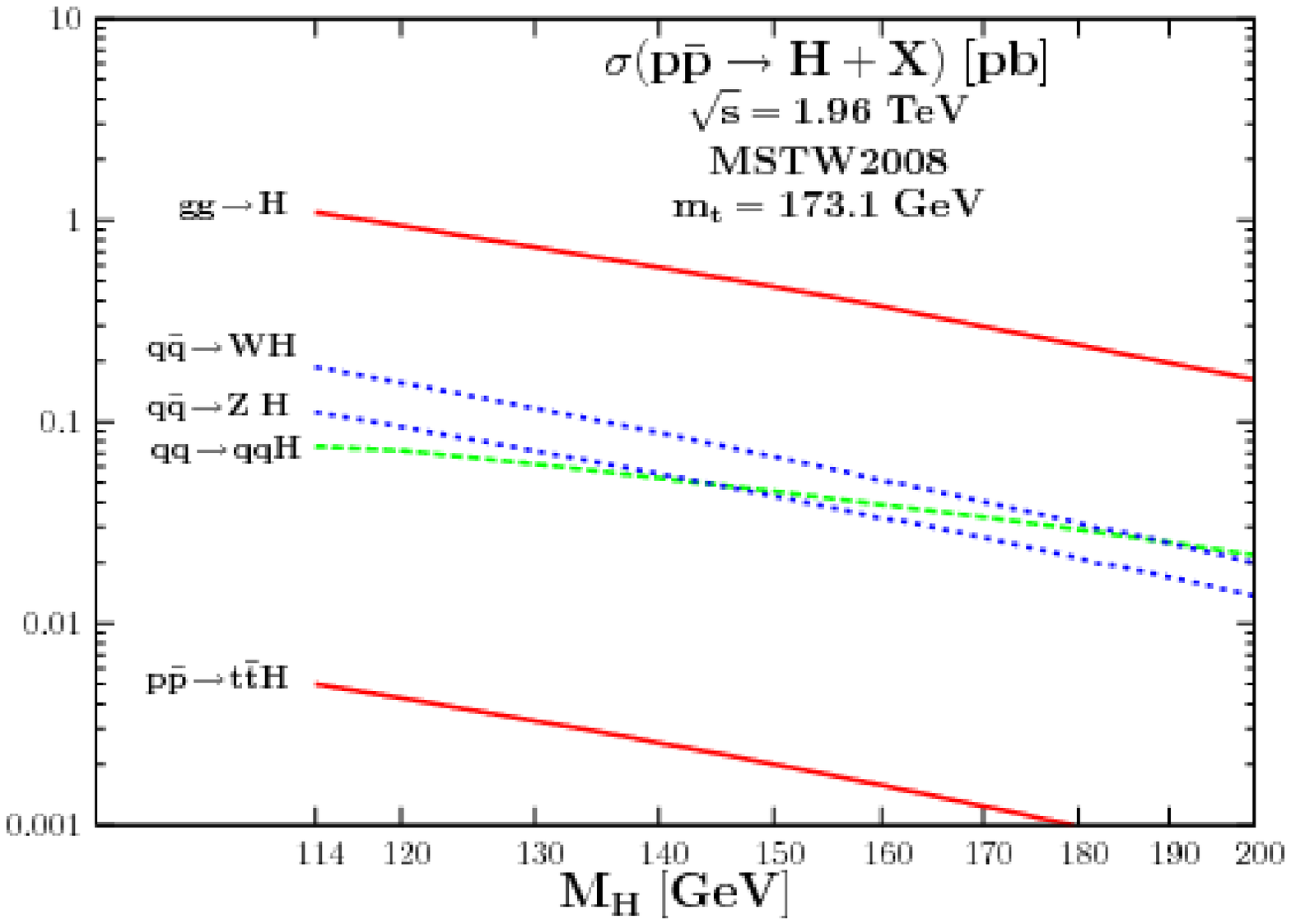}}
\end{minipage}
\hfill
\begin{minipage}{0.48\linewidth}
\centerline{
\includegraphics[width=0.75\linewidth]{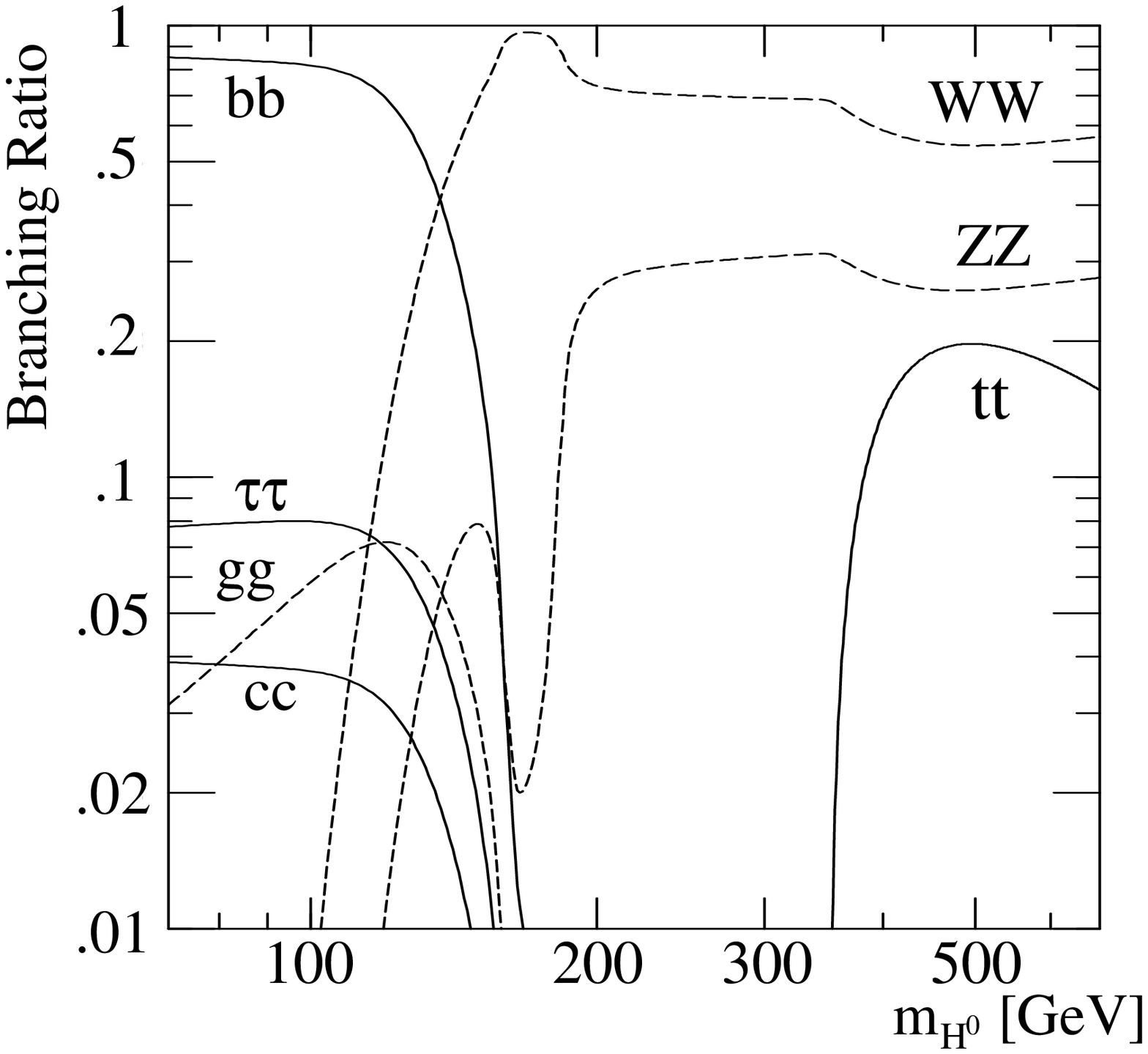}}
\end{minipage}
\hfill
}
\caption{SM Higgs production cross sections at the Tevatron (left) and its 
decay branching ratio (right) as a function of the Higgs boson mass.} 
\label{fig:xsec}
\end{figure}

 The dominant Higgs boson production processes at the Tevatron are gluon-gluon 
 fusion ($gg \rightarrow H$) and the associated production with a $W$ 
 or $Z$ boson~\cite{xsec}. Figure~\ref{fig:xsec} shows the Higgs boson 
 production cross sections at the Tevatron as a function of Higgs 
 boson mass and the Higgs decay branching ratio in the 
 standard model~\cite{hbr}. The Higgs boson decay is dominated by 
 $H\rightarrow b\bar b$ in the low-mass range ($m_H<130$ GeV/c$^2$) and by 
 $H\rightarrow W^+W^-$ or $Z Z^*$ in the high-mass range ($m_H>130$ GeV/c$^2$). 
 A search for a low-mass Higgs boson in the $gg\rightarrow H\rightarrow b\bar b$
 channel is extremely difficult because the $b\bar b$ QCD production cross 
 section is many orders of magnitude higher than the Higgs production. Requiring
 the leptonic decay of the associated $W$ or $Z$ boson greatly improves the 
 expected signal over background ratio in the $b\bar b$ channel, which makes it
 to be the most promising channel for the low-mass searches at the Tevatron.
 Other searches in the secondary channels for $H\rightarrow \gamma \gamma$, 
 $\tau^+\tau^-$, and $t\bar t H$ are also considered. We have to combine 
 all searches in many different channels in order to obtain the best 
 sensitivity. 

 The difficulty for the Higgs search at the Tevatron is due to the fact that 
 the Higgs signal is so small compared to other standard model processes with 
 the same final states. Search strategies employed by CDF and D0 have been 
 constantly improving over time. We first maximize the signal acceptances by 
 using efficient triggers, excellent lepton identifications, and b-tagging. 
 Then we use multivariate analysis (MVA) to exploit the kinematic differences 
 between signal and background, which can further enhance the 
 signal-to-background ratio from 1\% up to the 10\% level in the high score 
 regions. The procedures are iterated with improvements for every major 
 update until the best sensitivity is achieved. 

\begin{figure}
\centerline{
\includegraphics[width=0.3\linewidth]{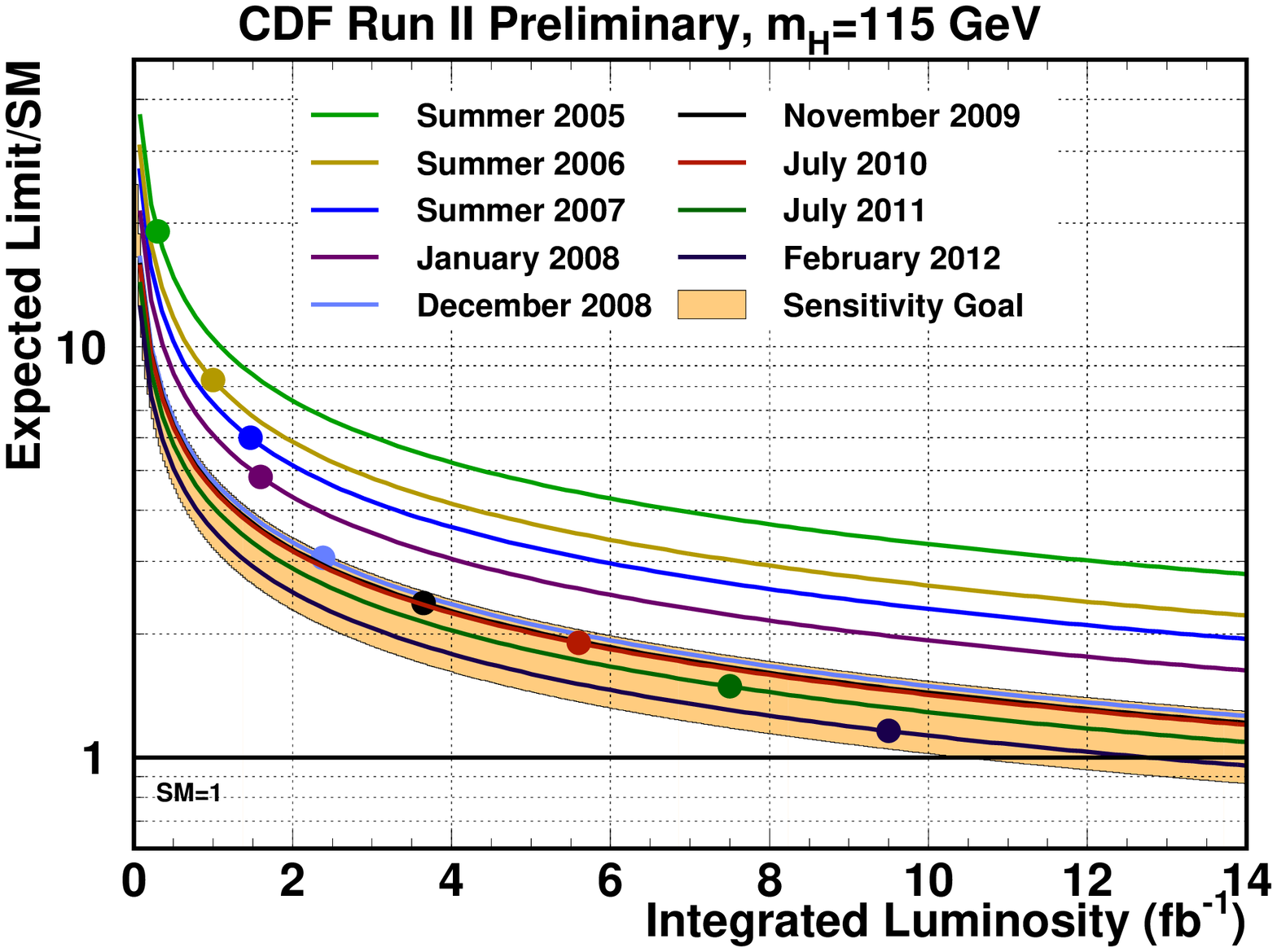}
\includegraphics[width=0.3\linewidth]{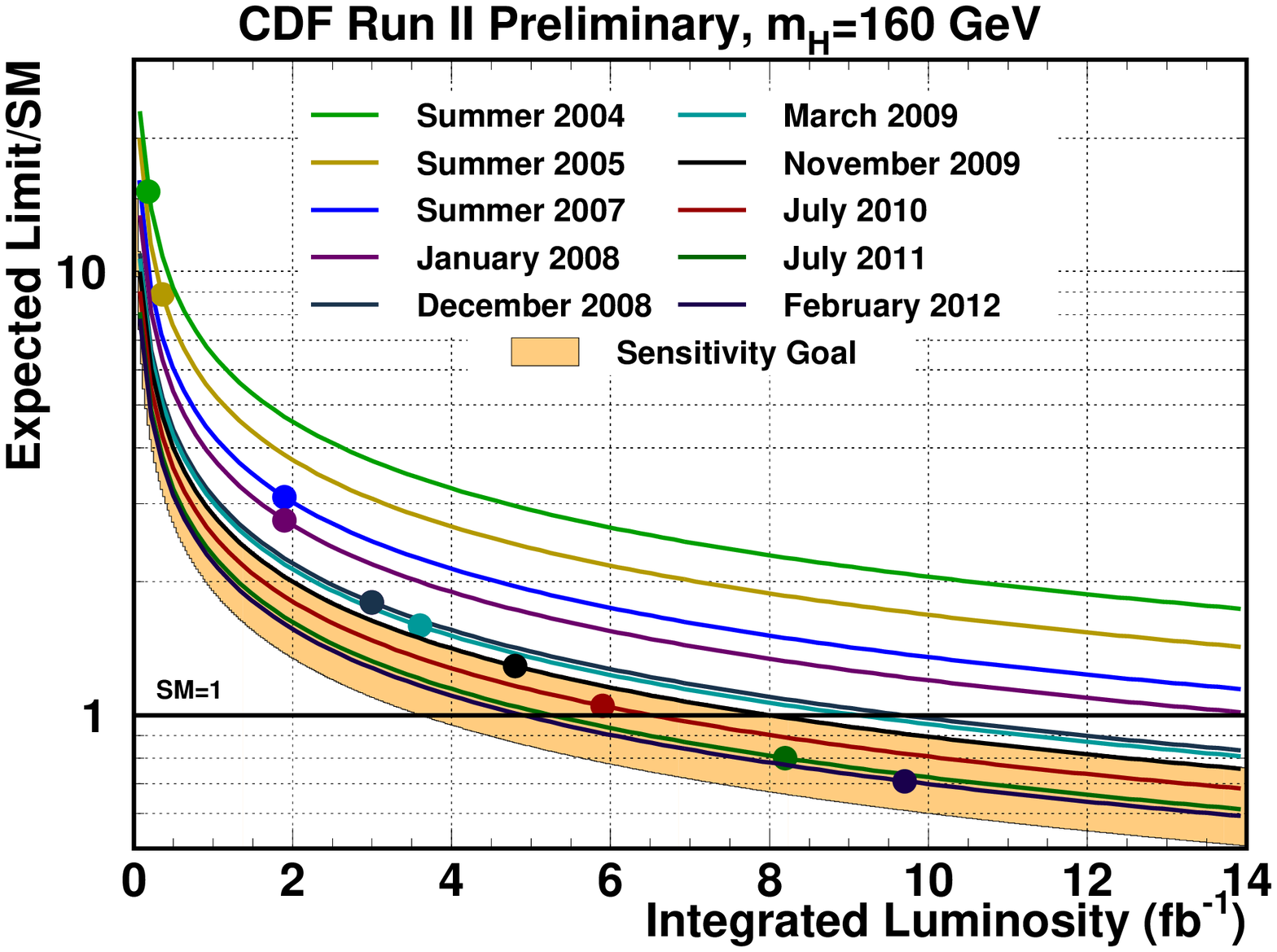}}
\hfill
\caption{Achieved and projected median expected upper limits on the SM 
Higgs boson cross section for $m_H$=115 GeV/c$^2$ (left) 
and 165 GeV/c$^2$ (right) as a function of integrated luminosity used for 
each major update. The solid lines are 1/sqrt(L) projections assuming no 
further improvements. The orange band corresponds to 
the Summer 2007 performance expected limit divided by 1.5 on top, and 
by an additional factor of 1.5 on bottom.}
\label{fig:tev115sensitivity}
\end{figure}

 Figure~\ref{fig:tev115sensitivity} shows the evolution of 
 achieved and projected median 
 expected upper limits on the SM Higgs boson cross section as a function of the 
 luminosity used for the low-mass searches (left) and the high-mass searches 
 (right). The solid lines are the luminosity projection assuming there is no 
 further improvement available. In the past, we have constantly introduced and
 improved analysis techniques to gain sensitivity far beyond 
 expectation from increased luminosity. The orange band corresponds to our 
 conservative and aggressive projections based on 2007 summer 
 results. The last update with full dataset indicates we have exceeded our 
 most optimistic expectations. 

\subsection{CDF and D0 Searches}

 Event selections are similar in the CDF and D0 analyses, typically consisting 
 of a preselection based on event topology and kinematics. Multivariate analysis
 (MVA) techniques are used to combine several discriminating variables into 
 a single final discriminant that is used in the analysis.
The gain of sensitivity from MVA is about 10-25\%, compared to use a 
 single best variable, such as the dijet mass. Each channel is then 
 divided into exclusive sub-channels according to various
 lepton, jet multiplicity, and different $b$-tagging categories. 
This procedure groups events with 
 similar signal-to-background ratio to optimize the overall sensitivity. 
 For $H\rightarrow b\bar b$ signatures we look for a $b\bar b$ mass resonance 
 in associated with a $W$ or $Z$ boson where the 
 $W$ decays into $l\nu$, the $Z$ decays into $l^+l^-$, and $\nu\bar \nu$. 
 The $WH\rightarrow l\nu b\bar b$ 
 channel is the most sensitive channel and we apply $b$-tagging and use 
 advanced multivariate analysis (MVA) to suppress the $W$+jets and 
 top backgrounds. 

  For high mass Higgs signatures, we look for Higgs boson decaying 
  into $W^+W^-$ pair in the inclusive $gg\rightarrow H$ events that lead to 
  many interesting final states.
  The most sensitive channel is both $W$ decaying leptonically that gives an  
  opposite-sign dilepton, a large missing transverse energy, and 
  plus some jets in the 
  final state. Due to the miss neutrinos, we can not reconstruct the Higgs mass 
  and have to rely on the event kinematic that distinguishes signal 
  from background. For example, the opening angle between two leptons in the 
  signal events is generally smaller than that in background events 
  due to the fact that the Higgs boson is a scalar particle. Furthermore, 
  the missing energy of momentum is larger and the total transverse 
  energy of the jets is lower than they are typically in background events.   
  We combine all these variables into a single final discriminant, which 
  improves the sensitivity significantly in the high-mass channel.
  
 Although the primary sensitivity for $m_H<130$ GeV/c$^2$
 is provided by the $H\rightarrow b\bar b$ analyses and for $m_H>130$ GeV/c$^2$
 by the $H\rightarrow W^+W^-$ analyses, additional sensitivity is achieved by 
 the inclusion of other channels. Both collaborations contribute analyses 
 searching for Higgs boson decaying into tri-lepton final states, $\tau$-lepton
 pairs, diphoton, and $t\bar tH$. 

\section{Combination Procedures} 
    We combine searches in many channels with different production cross
section and decay branching ratio, and
set limits with respect to nominal SM prediction. We combine the 
results using a combined likelihood formed from a product of likelihoods for 
the individual channels. Systematic uncertainties are treated as nuisance 
parameters with truncated Gaussian. The common systematic between CDF and D0 
are the luminosity, theoretical cross sections, and PDF uncertainties, which 
are treated as correlated. Other sources of systematic are experiment 
dependent, treated uncorrelated between experiments, but correlated within 
the experiment, such as lepton identification, $b$-tagging efficiency, 
jet-energy scale (JES), detector related efficiencies, and instrumented 
backgrounds. Most systematic parameters are constrained by the data in 
the background dominated regions. 

\begin{figure}
\centerline{
\includegraphics[width=0.3\linewidth]{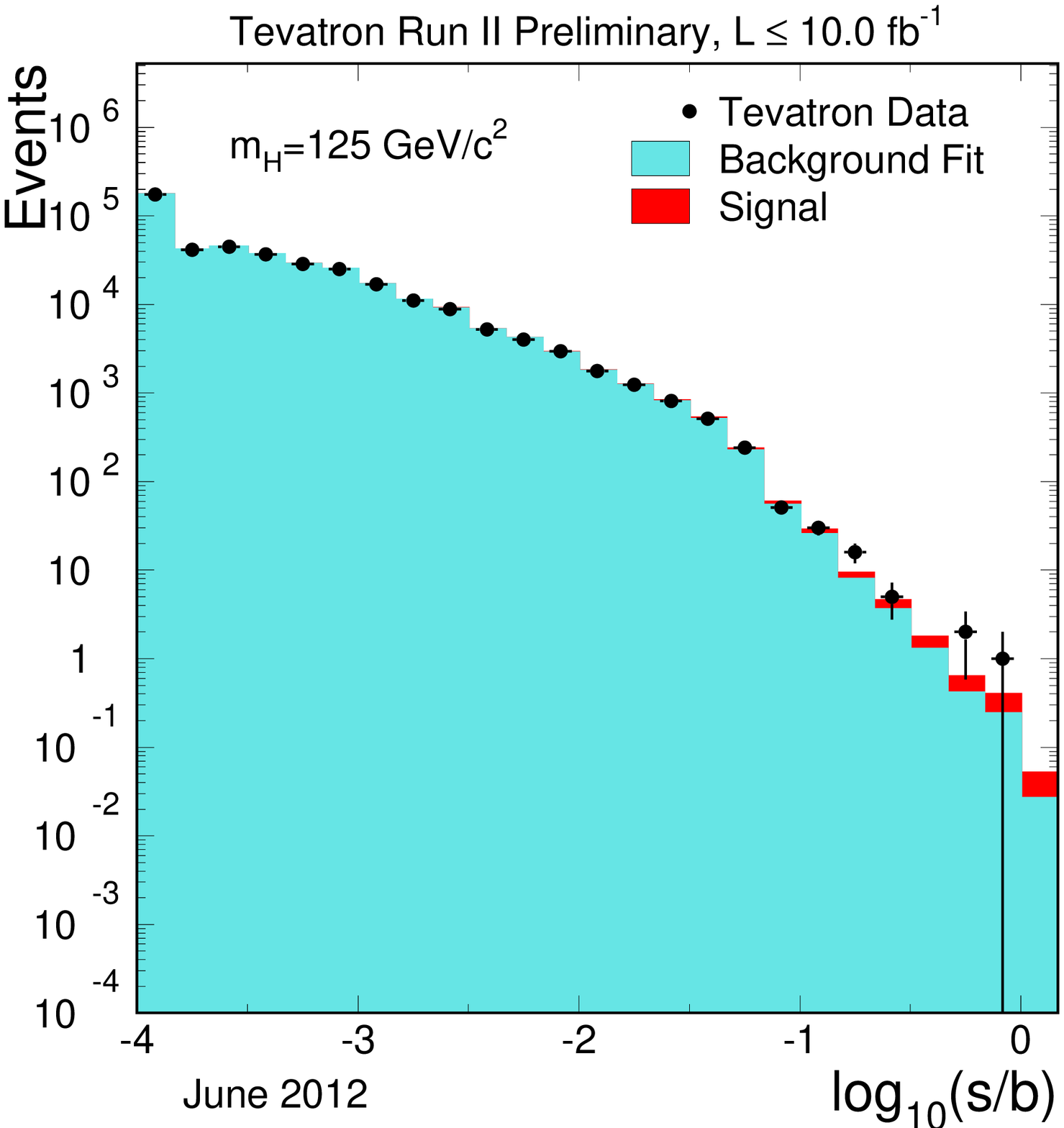}
\includegraphics[width=0.3\linewidth]{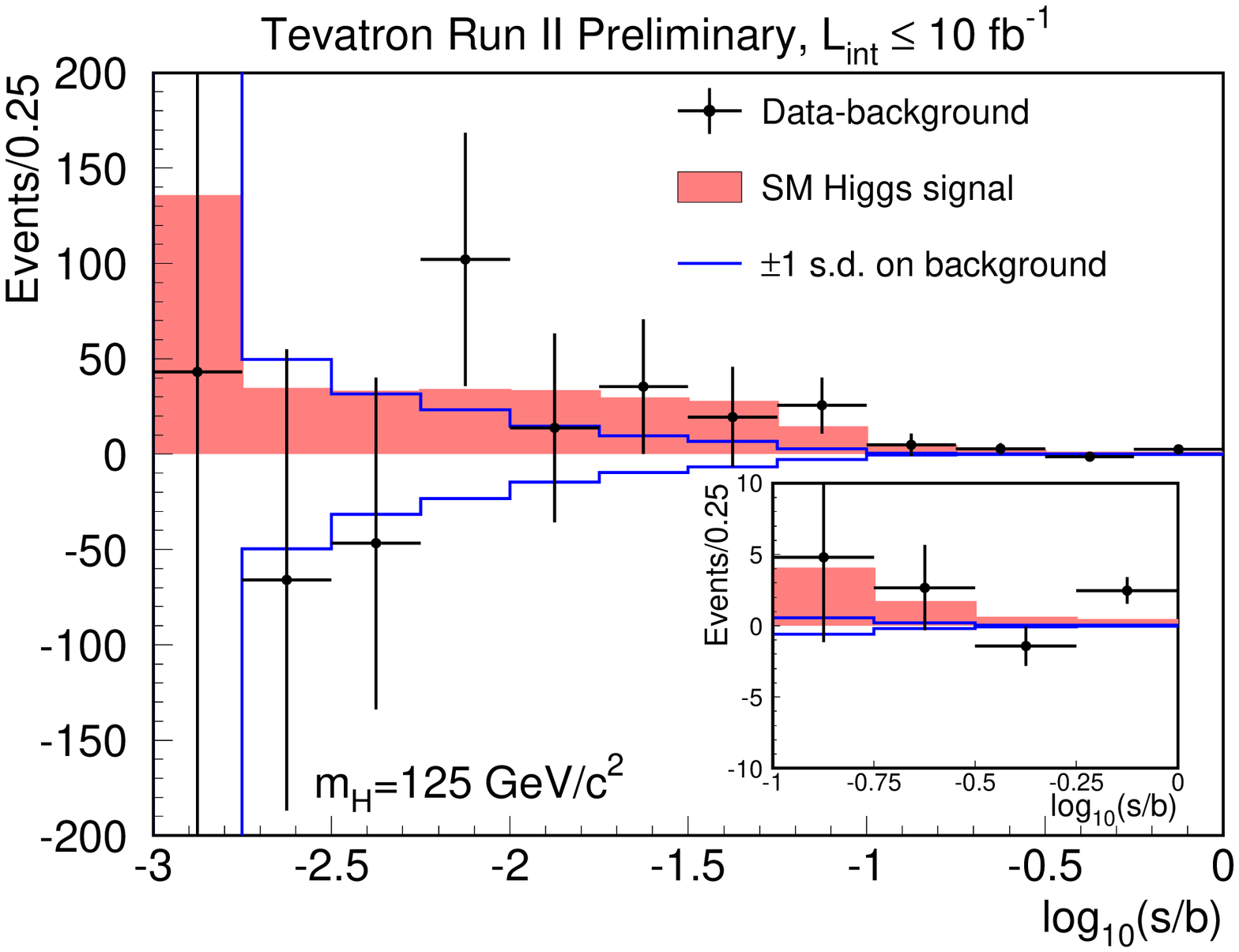}}
\hfill
\caption{Distribution of log$_{10}$(s/b) (left) for the data from all 
contributing Higgs boson search channels from CDF and D0 with 
$m_H$ = 125 GeV/c$^2$. The data are shown with points and the expected 
signal is shown red stacked on top of the background in blue. The corresponding
background-subtracted distribution is on right where the excess of events are 
consistent with the SM expectation of Higgs boson at $m_H$ = 125 GeV/c$^2$.}
\label{fig:tevsb125}
\end{figure}

In order to check the consistency between data and background, we rebin the
final discriminant from each channel in terms of signal-to-background ratio 
(s/b) so that the data with similar s/b may be added without loss of 
sensitivity. The resulting data distributions before and after background 
subtraction are shown in Figure~\ref{fig:tevsb125}.
There seems to be a small excess of Higgs boson
candidate events in the highest s/b bins relative to the background-only 
expectations, and the excess seems consistent with the expected signal of 
 $m_H=125$ GeV/c$^2$.  

\section{Diboson Searches} 
\begin{figure}
\centerline{
\includegraphics[width=0.3\linewidth]{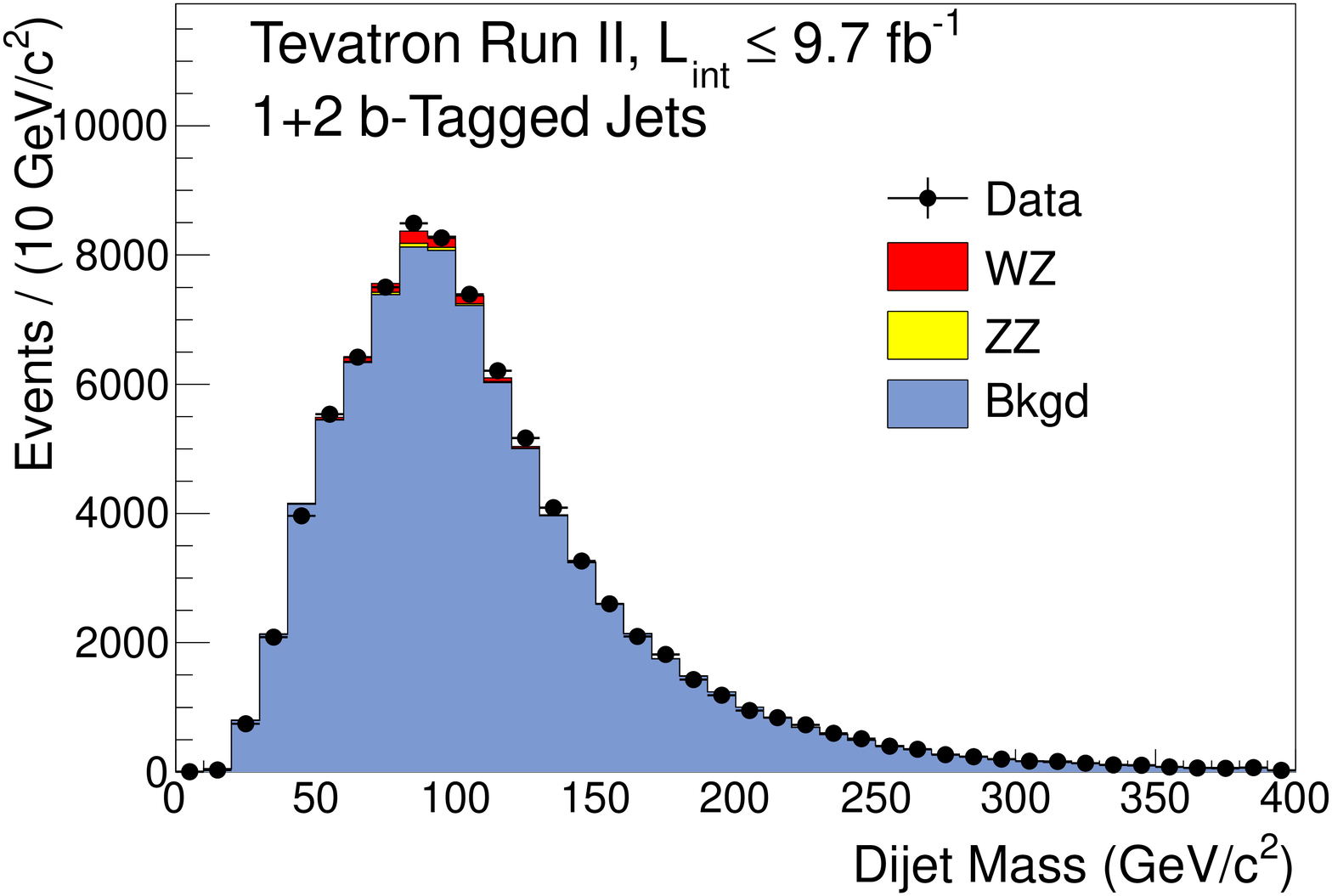}
\includegraphics[width=0.3\linewidth]{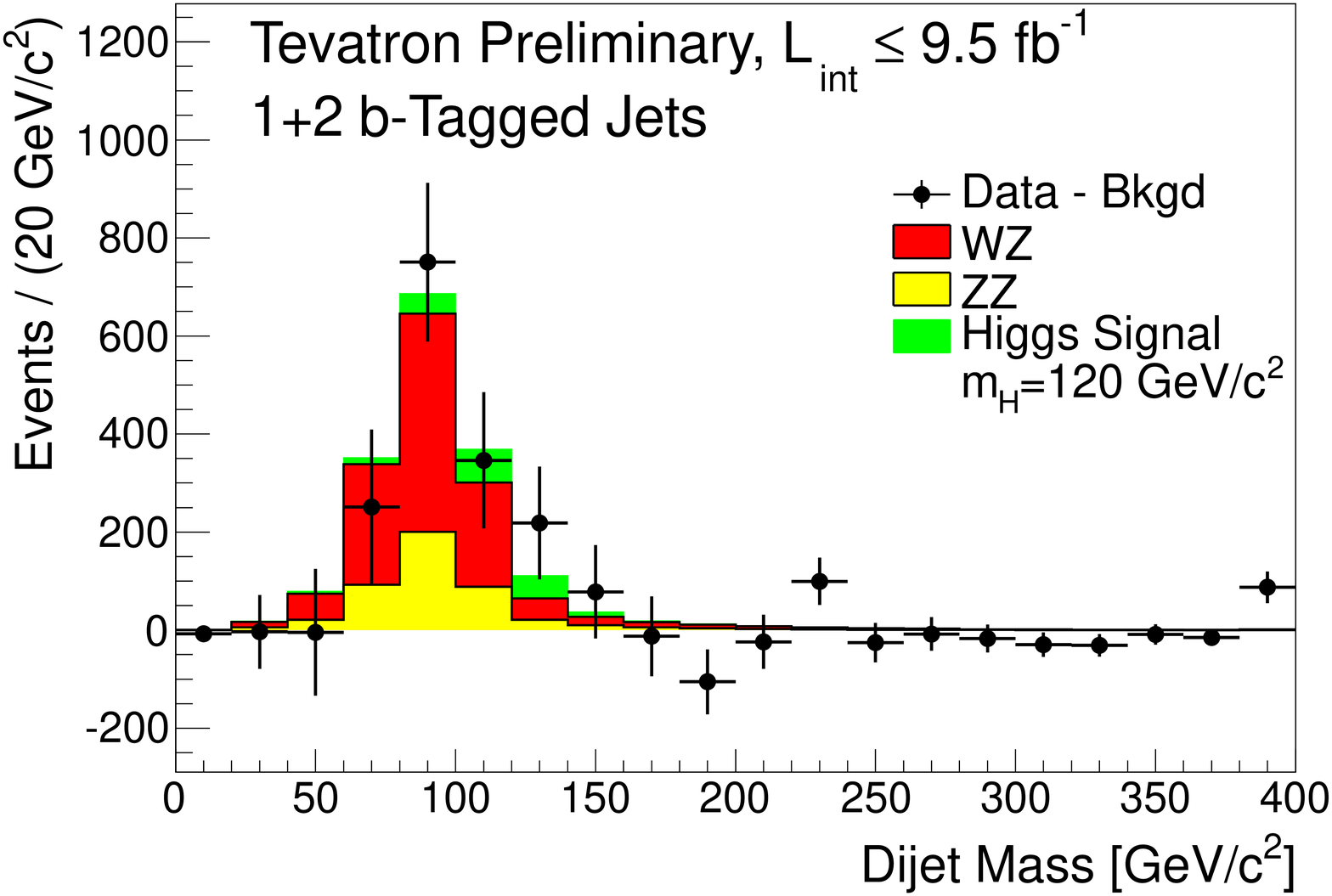}}
\hfill
\caption{The distribution of the reconstructed dijet mass, before (left) and
after (right) background-subtracted, summed over all CDF and D0's channels 
contributing to the VZ analysis. The fitted VZ and expected SM 
Higgs ($m_H$ = 125 GeV/c$^2$) contributions are also shown with 
filled histograms.}
\label{fig:VZmass}
\end{figure}

To validate our background modeling and search methods, we additionally 
perform a search for $Z\rightarrow b\bar b$ in association with a $W$ or $Z$ 
boson using the same final states of the SM $H\rightarrow b\bar b$ searches. 
The data sample, reconstruction, background models, uncertainties, and 
sub-channel divisions are identical to those of the SM Higgs boson search, 
but the discriminant functions are trained using the signal of 
$Z\rightarrow b\bar b$ instead. The measured cross sections of $WZ+ZZ$ is 
$3.0\pm 0.6\pm 0.7$ pb using the MVA discriminant, which is consistent with 
the SM prediction of $4.4\pm 0.3$ pb~\cite{VZxsec}. The combined 
dijet-mass distributions before and after background-subtracted 
for the $WZ+ZZ$ analysis is shown in 
Figure~\ref{fig:VZmass}. The signal and the background contributions are fit 
to the data, and the fitted background is then subtracted. The contribution 
expected from a SM Higgs signal with $m_H=125$ GeV/c$^2$ is also included.
The diboson cross sections measured by the high mass analysis are also in 
good agreement with SM predictions~\cite{wwd0}.  

\section{Combined Tevatron Searches} 

\begin{figure}
\centerline{
\includegraphics[width=0.3\linewidth]{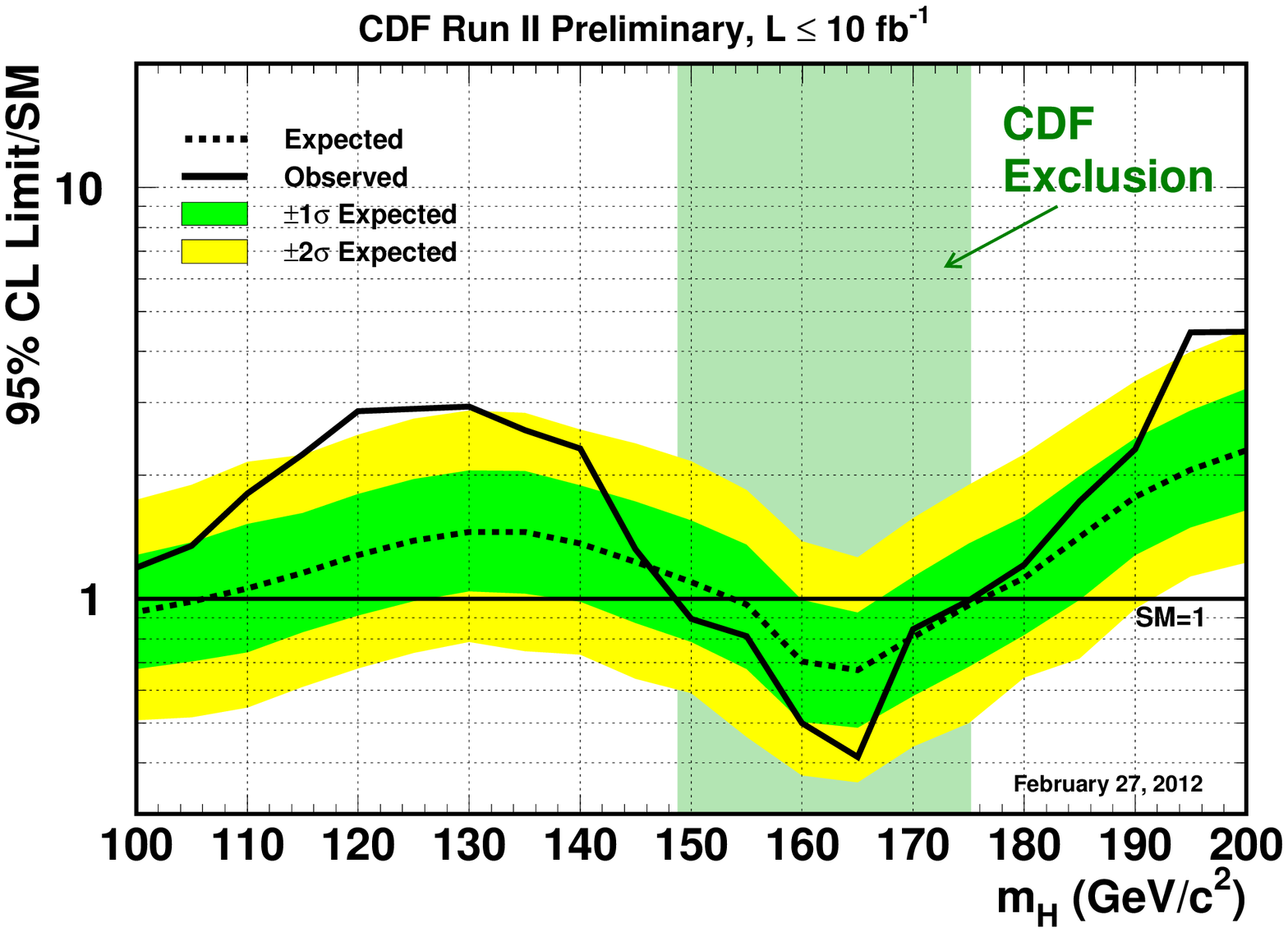}
\includegraphics[width=0.3\linewidth]{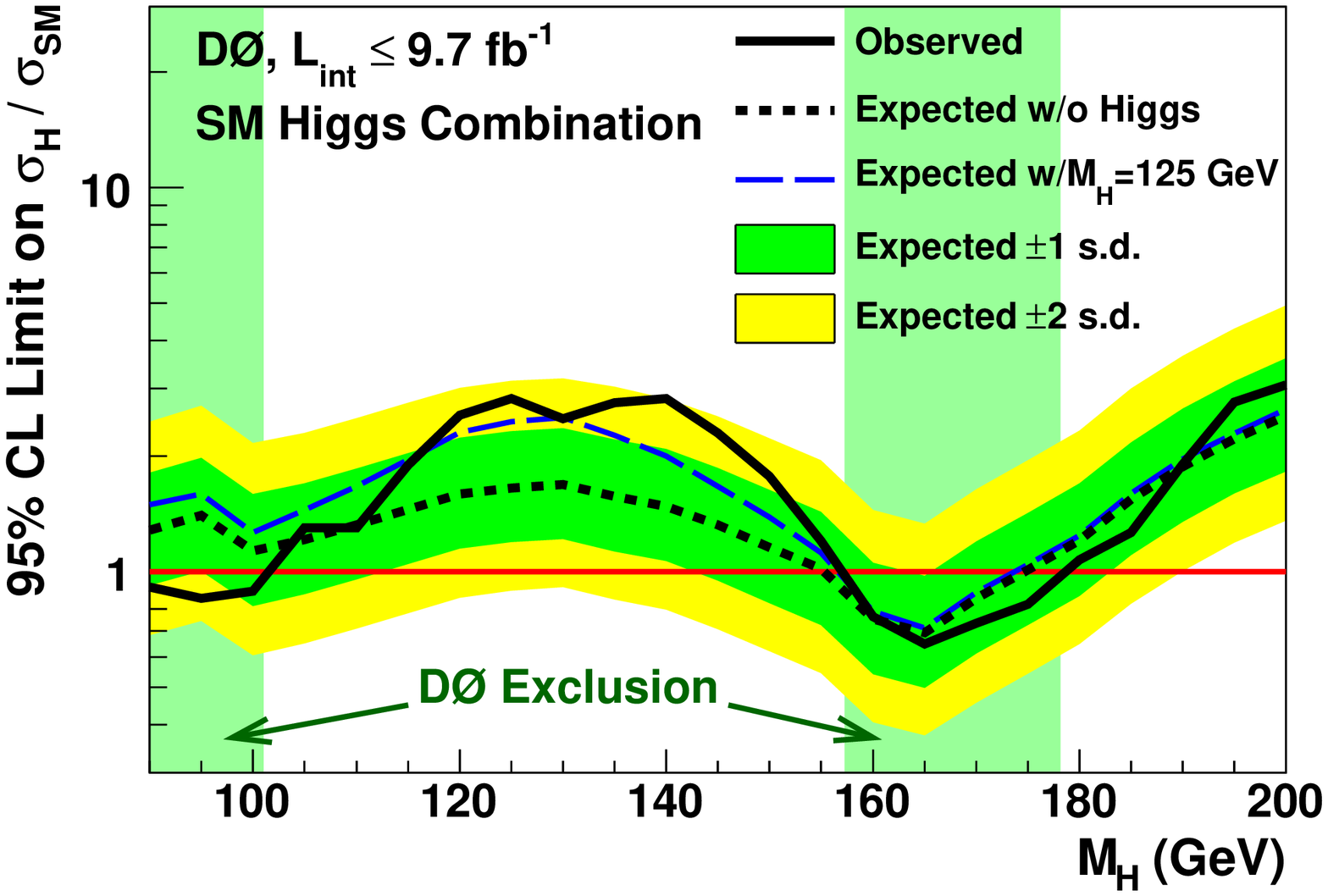}}
\hfill
\caption{The combined upper limit as a function of the Higgs boson mass 
between 100 and 200 GeV/c2 Solid black: observed limit/SM; Dashed black: 
median expected limit/SM. Colored bands: +-1, 2 sigma distributions around 
median expected limit from each experiment: CDF (left) and D0 (right).} 
\label{fig:individuallimits}
\end{figure}

There are no major changes in the analyses since the last update reported at 
the HCP conference last November~\cite{hcp2012}. Both collaboration have 
done many cross checks and are focusing on to get the final analyses out for
publication. Figure~\ref{fig:individuallimits} shows the combined limit from 
each experiment. They have very similar search sensitivities and comparable 
results. 

\begin{figure}
\begin{minipage}{0.33\linewidth}
\centerline{
\includegraphics[width=0.9\linewidth]{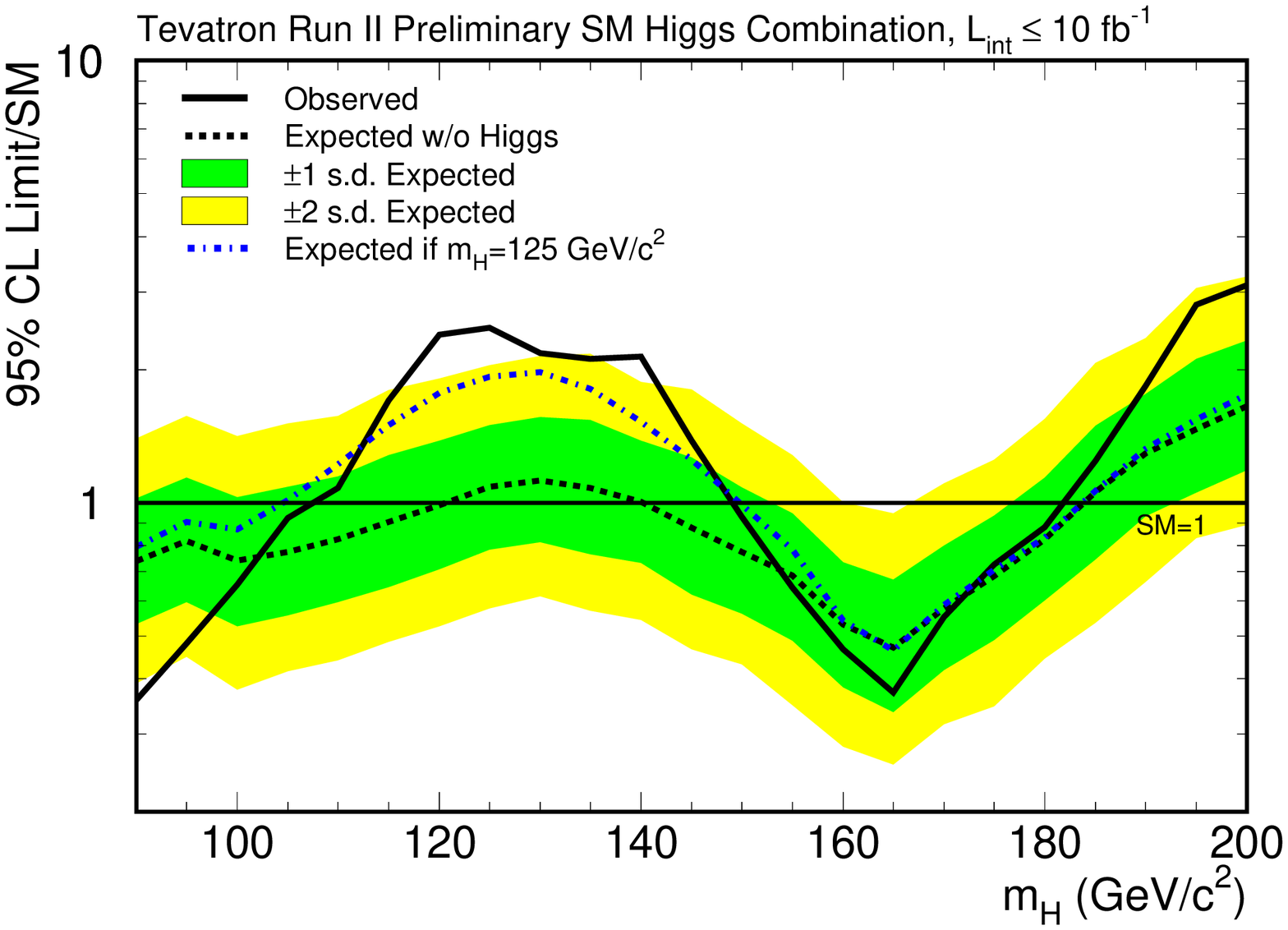}}
\end{minipage}
\hfill
\begin{minipage}{0.32\linewidth}
\centerline{
\includegraphics[width=0.9\linewidth]{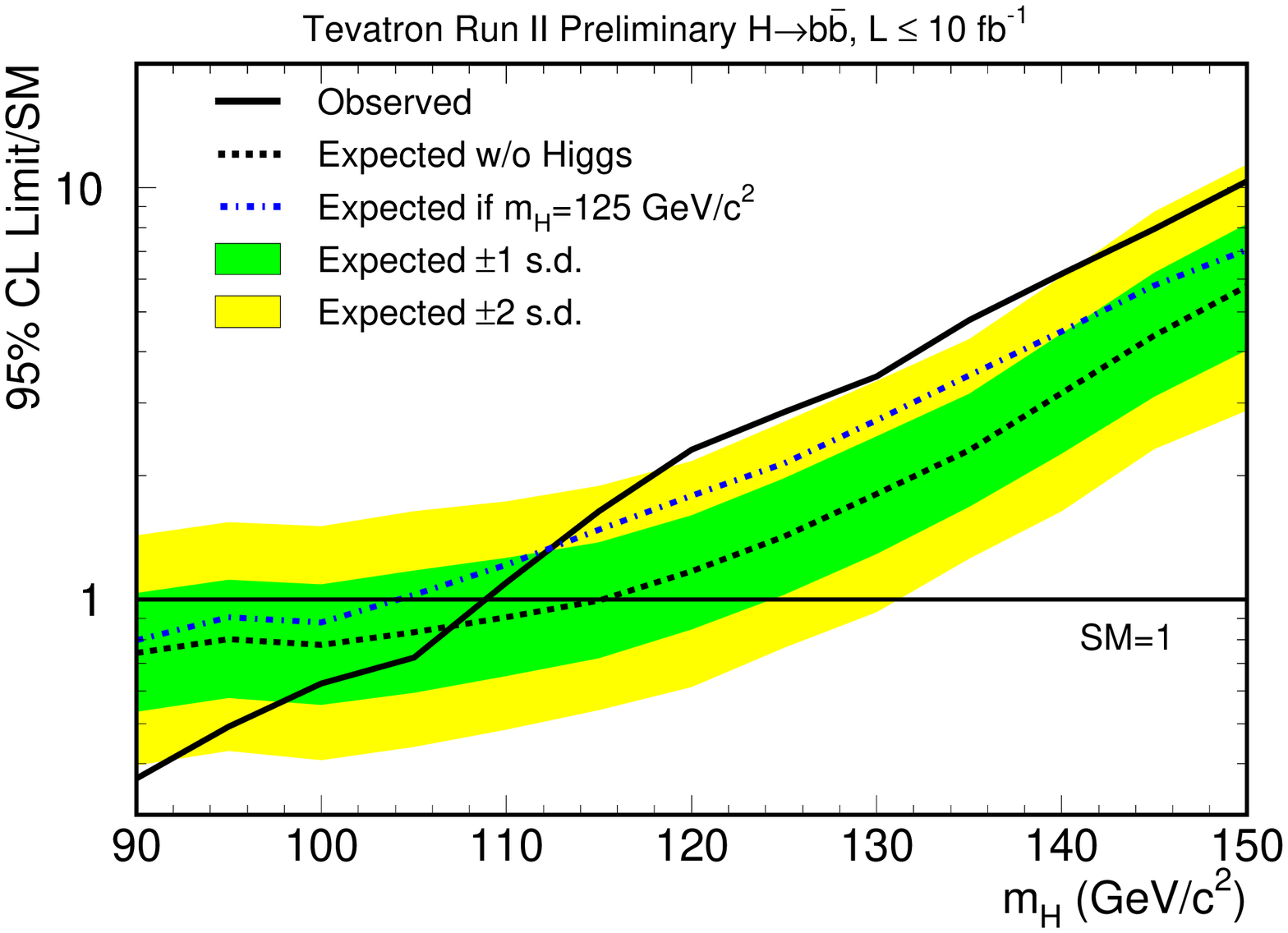}}
\end{minipage}
\hfill
\begin{minipage}{0.32\linewidth}
\centerline{
\includegraphics[width=0.9\linewidth]{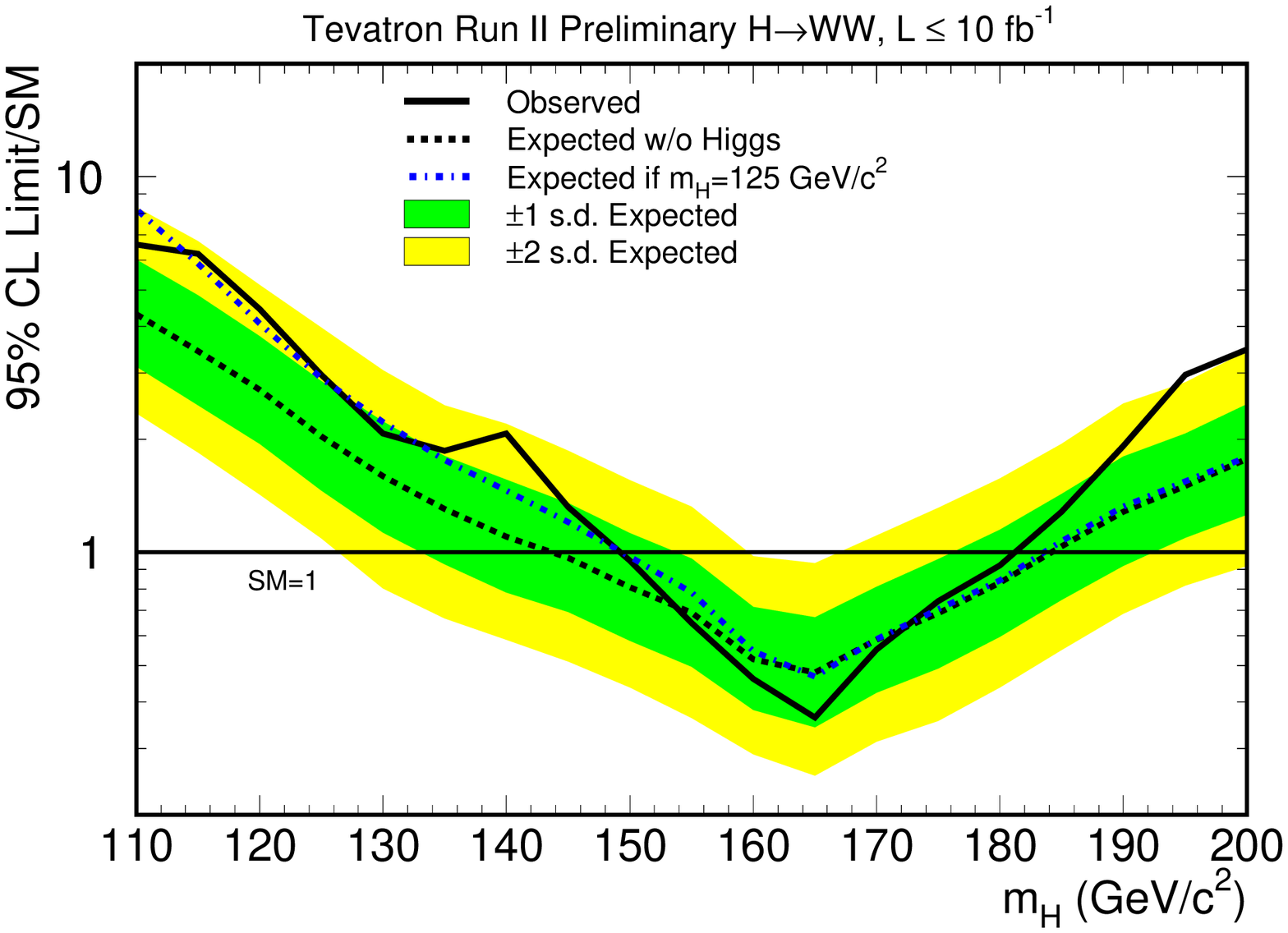}}
\end{minipage}
\hfill
\caption{Observed and median expected 95\% C.L. upper limits on the 
production times branching ratio respect to the SM prediction as a function of 
Higgs boson mass for the combined CDF and D0 searches in all decay channels 
(left), in the $b\bar b$ channels only (center), and in the $W^+W^-$ channels 
only (right), respectively. The dark- and light- shaded bands indicate the one 
and two standard deviation probability regions in which the limits are expected
to fluctuate in the absence of signal. The blue short-dashed line shows median
expected limits assuming the SM Higgs boson is present 
at $m_H$ = 125 GeV/c$^2$.}
\label{fig:tevlimits}
\end{figure}

\begin{figure}
\centerline{
\includegraphics[width=0.3\linewidth]{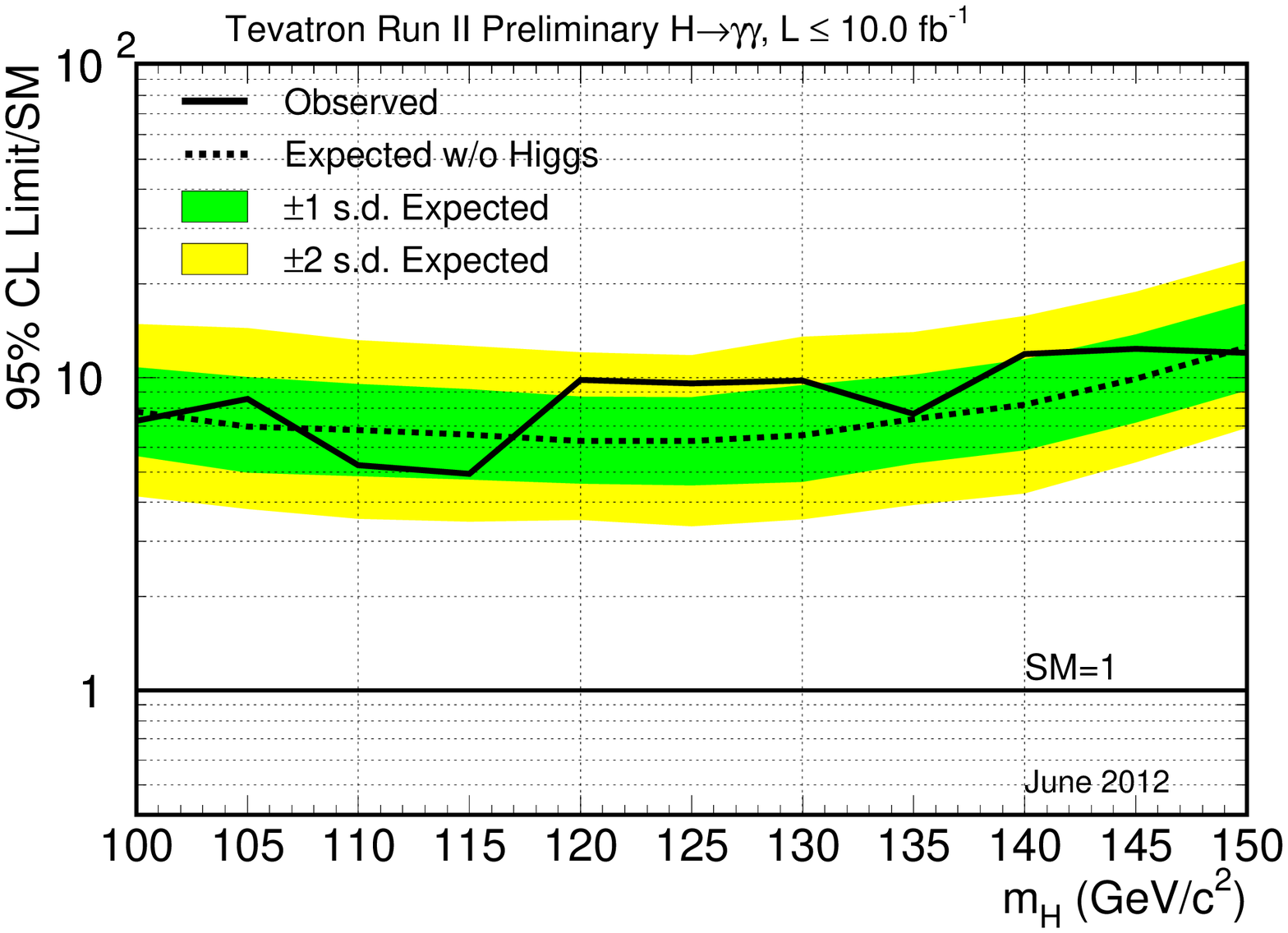}
\includegraphics[width=0.3\linewidth]{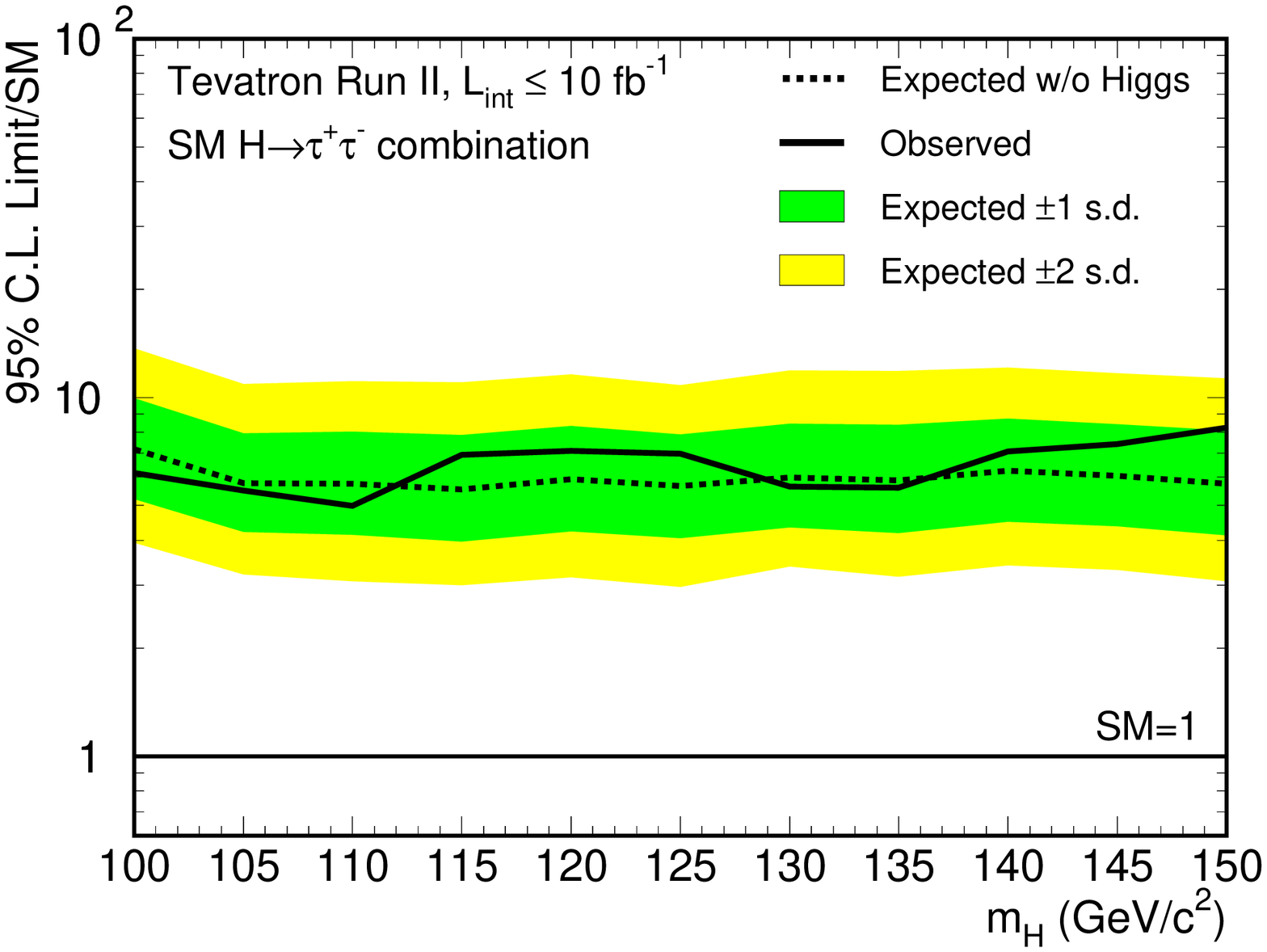}}
\hfill
\caption{Observed and median expected 95\% C.L. upper limits on the 
production times branching ratio respect to the SM prediction as a function of 
Higgs boson mass for the combined CDF and D0 searches in the 
$H\rightarrow \gamma\gamma$ channels 
(left) and in the $H\rightarrow\tau^+\tau^-$ channels (right), respectively. 
The dark- and light- shaded bands indicate the one 
and two standard deviation probability regions in which the limits are expected
to fluctuate in the absence of signal.}
\label{fig:tevsecondary}
\end{figure}

The ratio of the 95\% CL limit on SM Higgs boson production to  
the SM expected rate is extracted as a function of $m_H$ in the range 
90-200 GeV/c$^2$, as shown in 
Figure~\ref{fig:tevlimits} after combining the CDF and D0 analyses together. 
We are able to exclude the regions of Higgs boson masses at 95\% CL in 
90-107 GeV/c$^2$ and 149-182 GeV/c$^2$. The expected exclusion regions are 
90-121 GeV/c$^2$ and 140-184 GeV/c$^2$. There is a broad  excess observed of 
events between 115-140 GeV/c$^2$. We also separate CDF and D0's searches into 
combination limits on the $H\rightarrow b\bar b$, $H\rightarrow W^+W^-$, 
$H\rightarrow \gamma\gamma$, and $H\rightarrow \tau^+\tau^-$ decay modes, 
as shown in Figure~\ref{fig:tevlimits} and Figure~\ref{fig:tevsecondary}.
   
\begin{figure}
\centerline{
\includegraphics[width=0.3\linewidth]{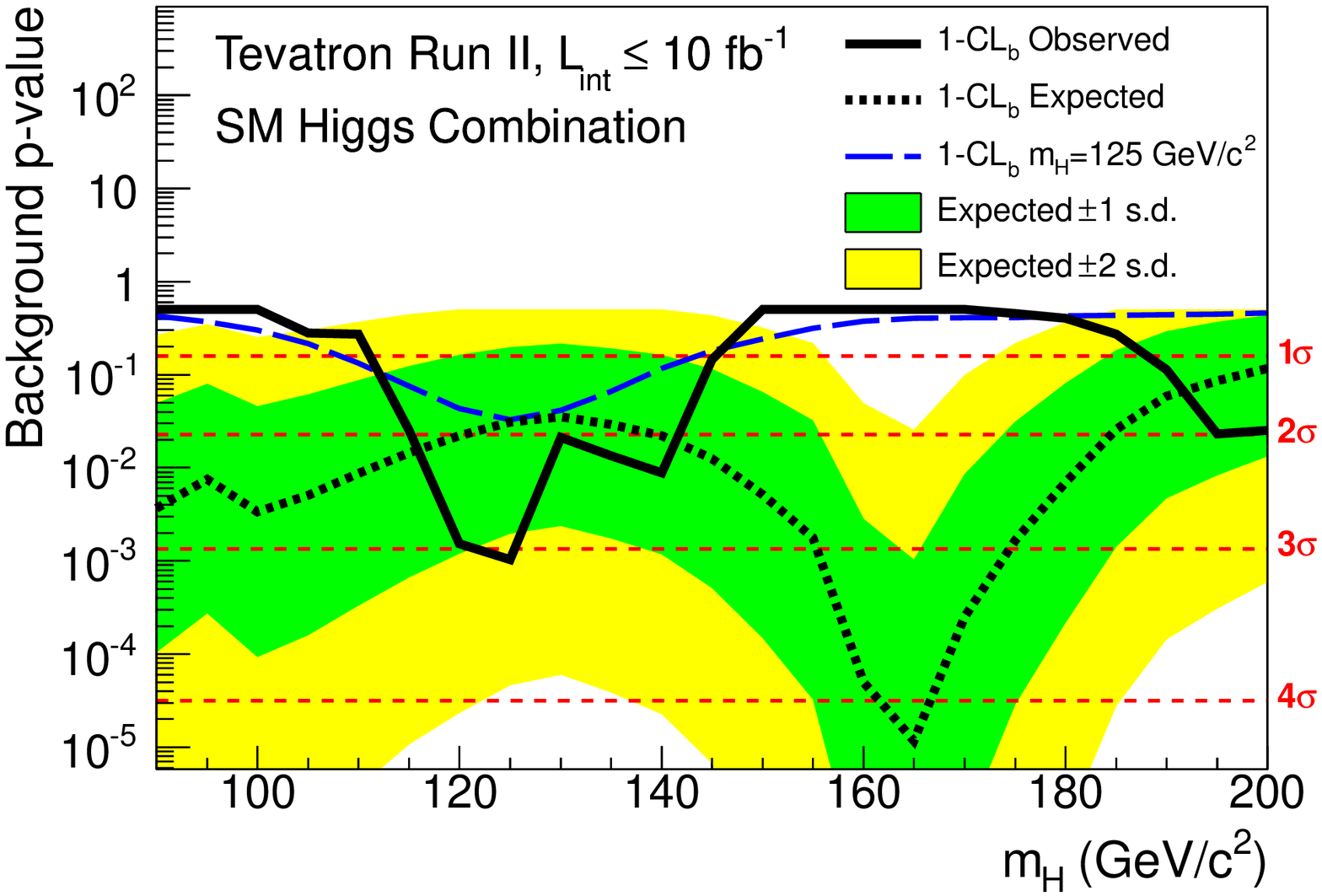}
\includegraphics[width=0.3\linewidth]{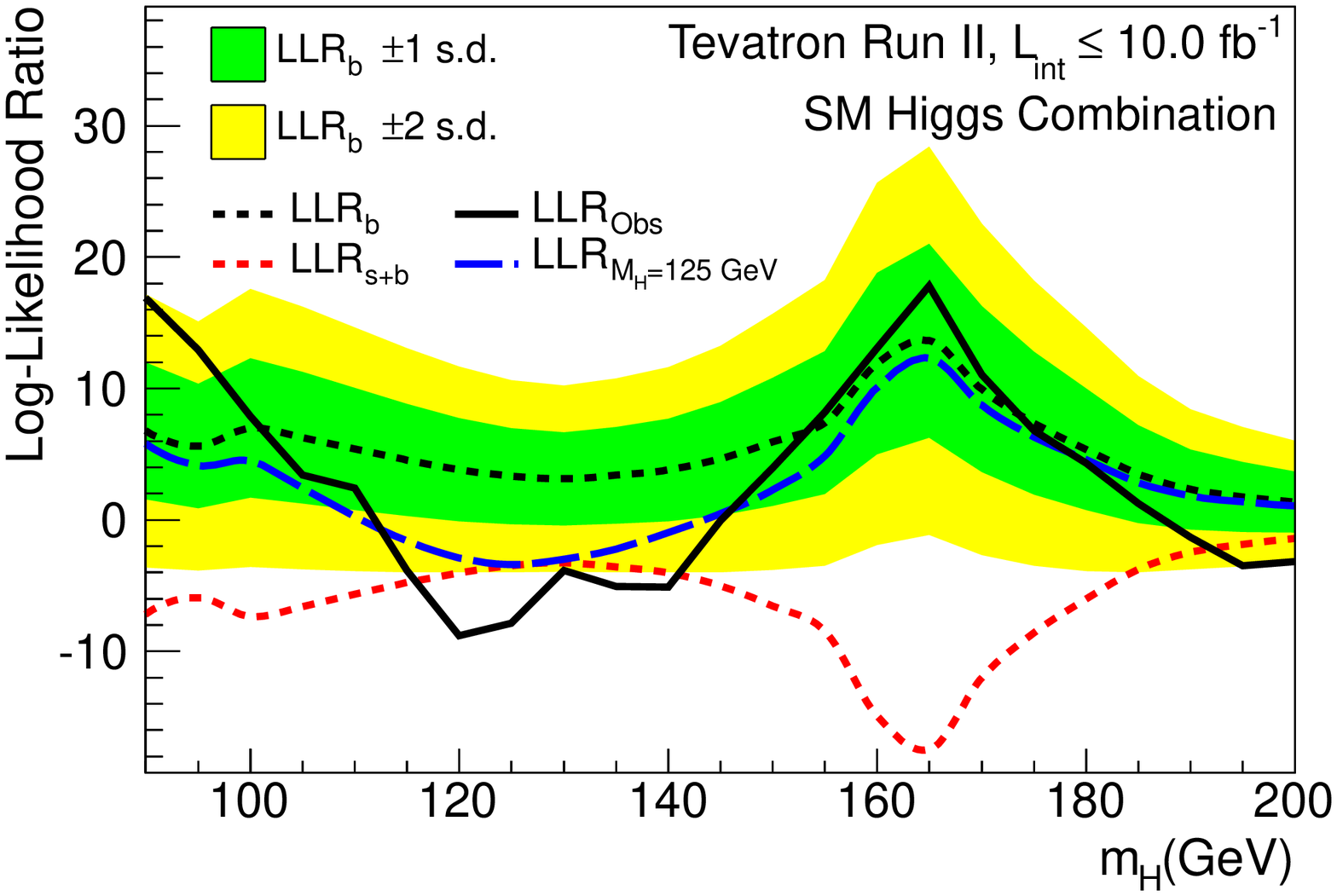}}
\hfill
\caption{The local background p-value (left) and the log-likelihood ratio
LLR (right) are shown as a function of Higgs boson mass for the combined CDF 
and D0 searches in all channels. The solid lines shows the observed values, 
the dark short-dash lines show the expected value, and  the dark- and light- 
shaded bands indicate the one and two standard deviation probability 
regions in which the values are expected to fluctuate.}
\label{fig:pvalues}
\end{figure}

  We quantify the excess by calculating the local p-value for the 
background-only hypothesis. Figure~\ref{fig:pvalues}
shows the local p-value 
as a function of Higgs boson mass where the solid curve is for data and the 
dash curve is what expected from the SM Higgs boson. At $m_H$=125 GeV/c$^2$, we 
have a minimum local p-value corresponding to 3 $\sigma$ standard deviation. 
The broken line is what expected if a Higgs signal of 125 GeV/c$^2$ is present,
which has similar shape to the data.  

 To further check the compatible with the SM Higgs signal 
 of $m_H$=125 GeV/c$^2$, 
 we compared the log-likelihood-ratio (LLR) 
 after injecting $m_H$=125 GeV/c$^2$ signal into background-only
  pesudo-experiment, as shown in broken line in Figure~\ref{fig:pvalues}. 
 The shape seems very similar to what observed in the data and 
 is quite broad due to the fact that MVA is not optimized for
 mass, but for signal-to-background separation.

\begin{figure}
\centerline{
\includegraphics[width=0.3\linewidth]{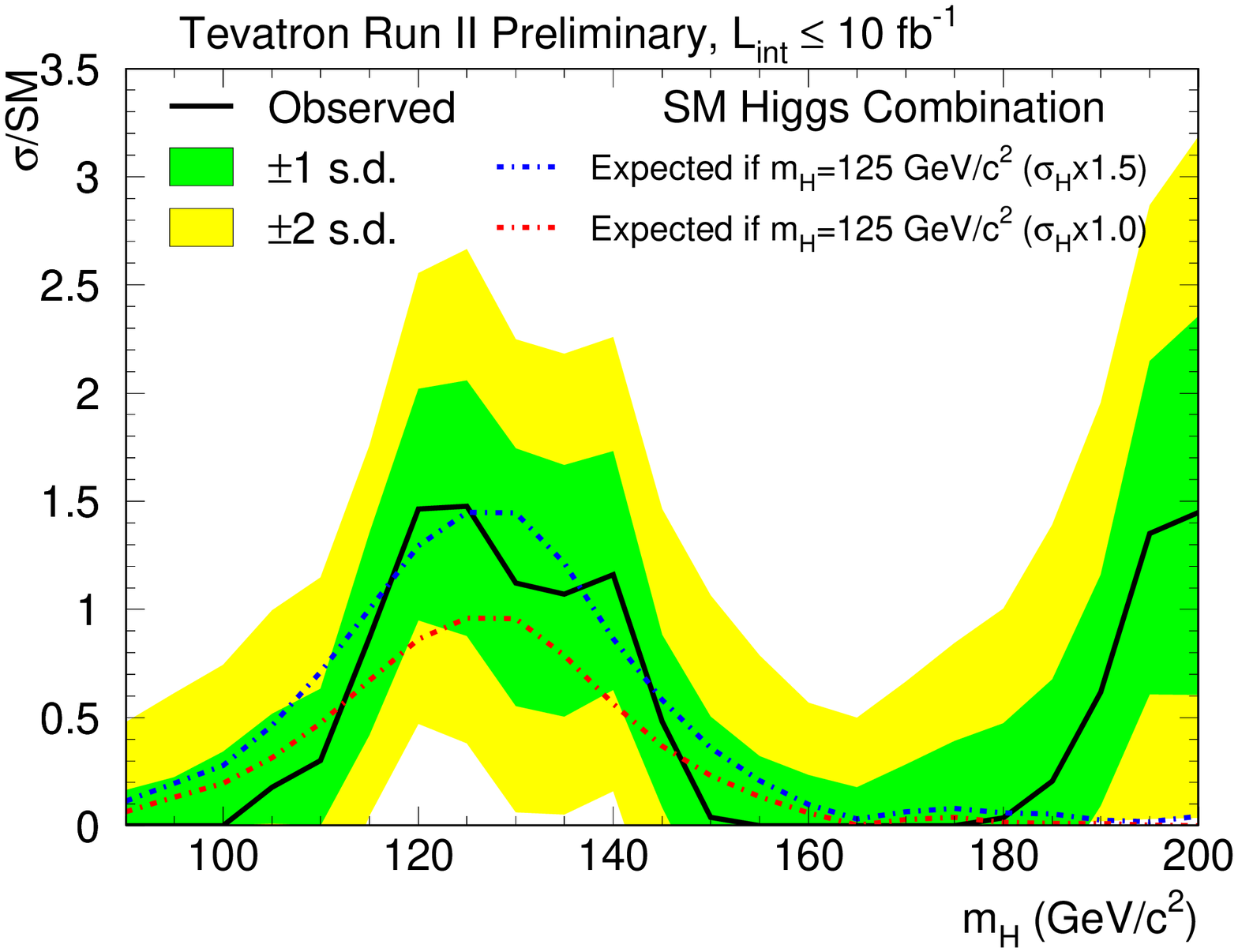}
\includegraphics[width=0.3\linewidth]{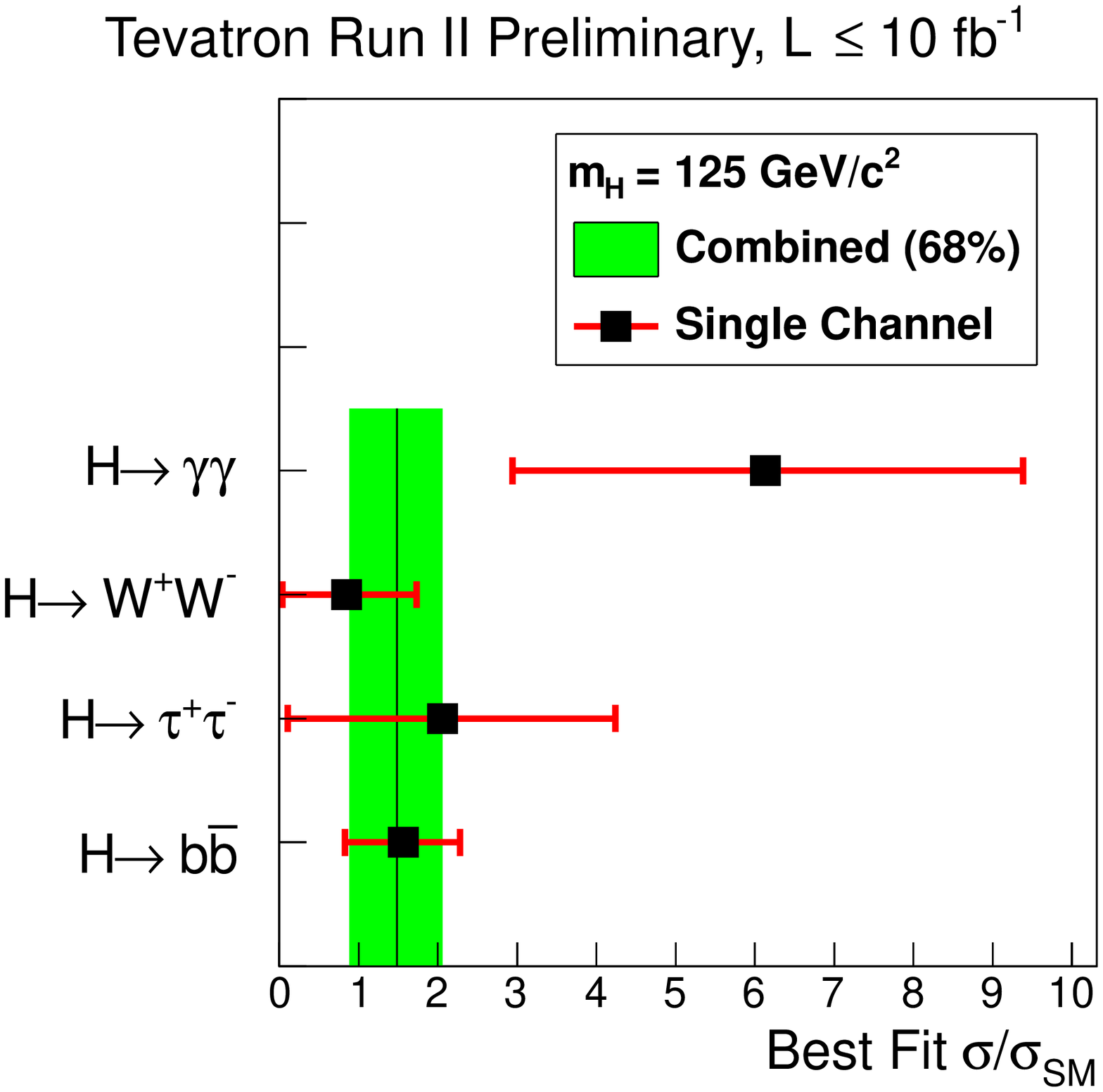}}
\hfill
\caption{The best-fit signal strength as a function of Higgs boson mass for 
the combined CDF and D0 searches in all channels (right) and in
separate channels for $H\rightarrow\gamma\gamma$, 
$W^+W^-$, $\tau^+\tau^-$,  and $b\bar b$ for a Higgs boson 
mass of 125 GeV/c$^2$ (right), respectively. 
The dark (light) shaded band corresponds to one (two) standard deviation 
uncertainty on the best-fit value of signal strength with all the channels
combined. Also shown on left with blue lines are the median fitted 
cross sections expected for a SM Higgs boson of $m_H$=125 GeV/c$^2$ with 
signal strengths of 1 (short-dashed) and 1.5 (long-dashed) times 
the SM prediction. }
\label{fig:signalstrength}
\end{figure}

 The observed excess of $m_H$ from 115 to 140 GeV/c$^2$ is driven by an excess
of data events with respect to the background predictions in the most 
sensitive bins of the discriminant distributions. We fit this excess with the 
signal-plus-background hypothesis and obtain the best-fit signal strength as 
a function of $m_H$, shown in Figure~\ref{fig:signalstrength}. 
The data are 
consistent with the expectation of a SM Higgs boson in the mass range 
115-140 GeV/c$^2$.  At $m_H=125$ GeV/c$^2$, the measured signal strength is 
 $1.44^{+0.59}_{-0.56}$. 

We also fit to data in separate channels for $H\rightarrow\gamma\gamma$, 
$W^+W^-$, $\tau^+\tau^-$,  and $b\bar b$ for a Higgs mass of 125 GeV/c$^2$, 
as shown in Figure~\ref{fig:signalstrength}.
The results are consistent with each other, with the full combination, and 
with the production of the SM Higgs boson as well. It's worth noting that 
the best-fit rate cross section for $(\sigma_{WH}+\sigma_{ZH})\times 
B(H\rightarrow b\bar b)$ has measured to be $0.19^{+0.08}_{-0.09}$ pb 
for $m_H=125$ GeV/c$^2$ , which is
slightly lower, but consistent with the value of $0.23\pm 0.09$ pb obtained 
previously~\cite{tevbb} due to the updated $ZH\rightarrow \nu\bar\nu b\bar b$ 
analysis from CDF~\cite{nunubb}. 

\section{Higgs Boson Properties}
\begin{figure}
\centerline{
\includegraphics[width=0.45\linewidth]{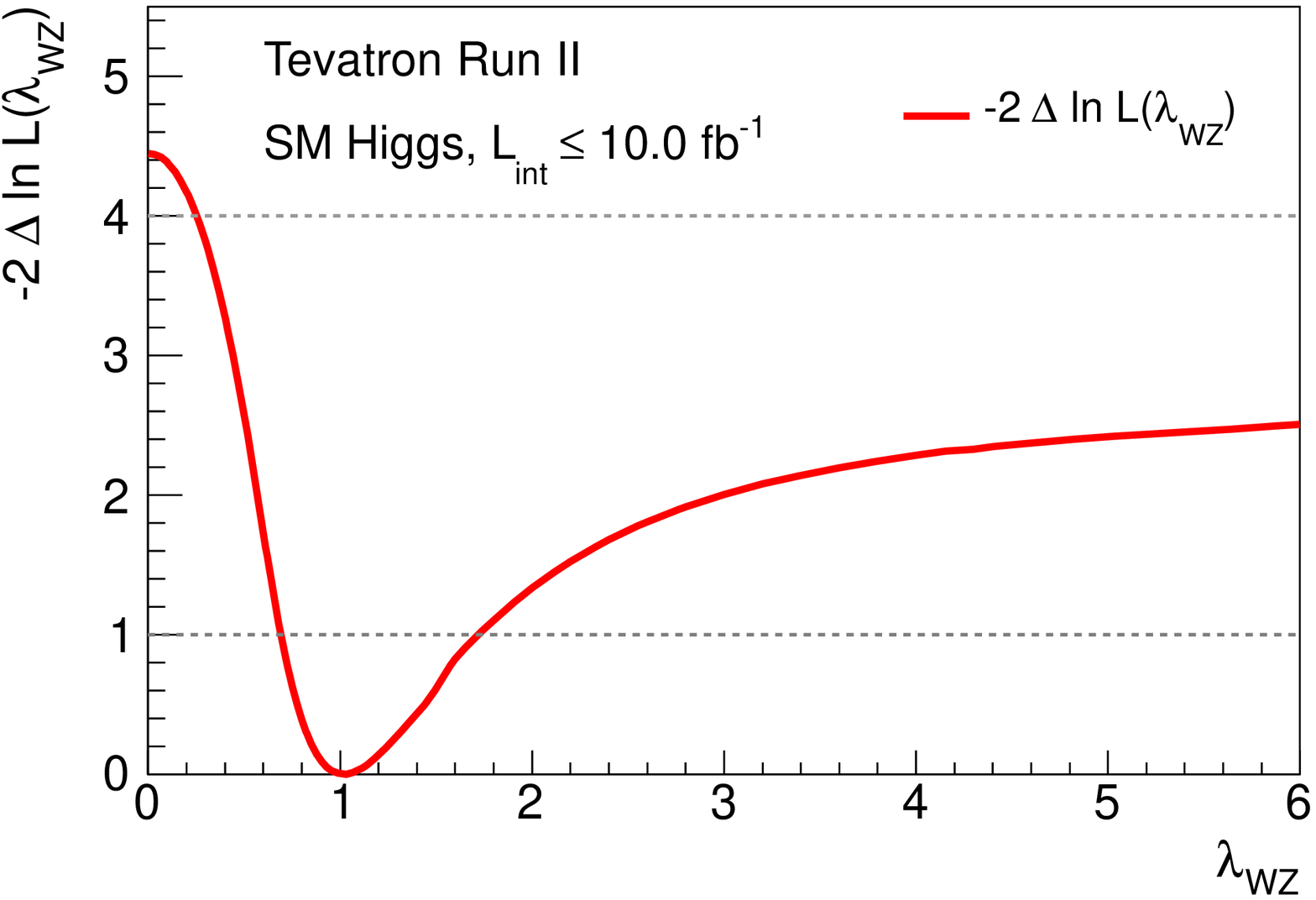}
\includegraphics[width=0.3\linewidth]{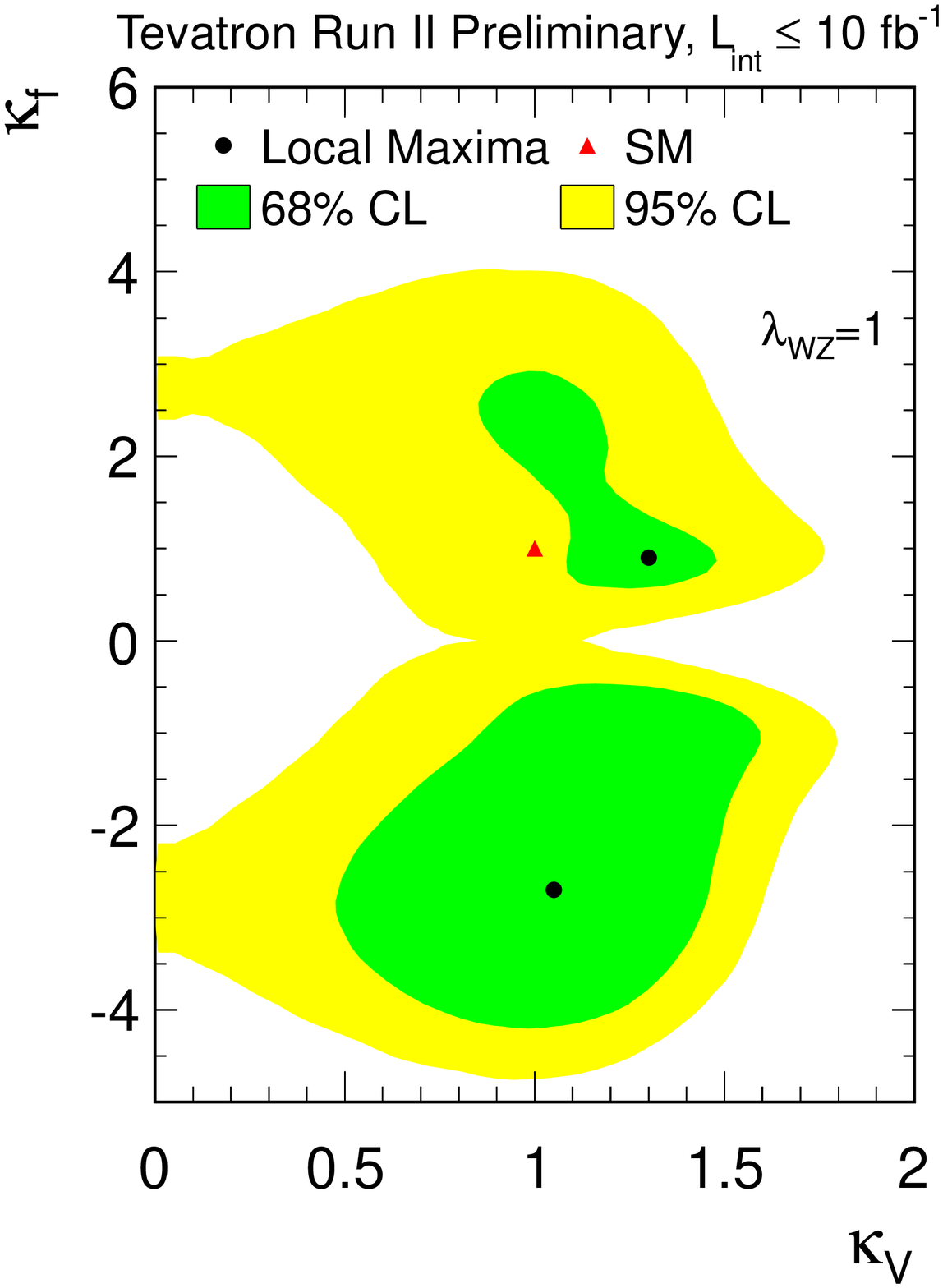}}
\hfill
\caption{The constraining of the custodial symmetry 
$\lambda_{WZ}$, the ratio of Higgs coupling to $W$ boson and $Z$ boson, 
by fixing all fermion coupling to SM predictions  for the combined Tevatron 
searches for a Higgs mass of 125 geV/c$^2$ (left) and  
two-dimensional constraining of fermion and boson coupling simultaneously by 
fixing $\lambda_{WZ}$ = 1 (right), respectively. The points on right show 
the best fit from the data with dots, and the dark- and light-shaded regions 
corresponding to the 68\% and 95\% CL intervals. The SM prediction is marked 
with a triangle. }
\label{fig:coupling}
\end{figure}

 In SM, the Higgs boson coupling to other particles is  proportional to 
 their mass. Probing its coupling may provide sensitivity to understand what
 the new particle is, which can be parametrized through coupling factors 
 respect to SM predictions. Most searches at the Tevatron are sensitive to 
 the product of fermion and boson coupling strengths. In the 
$VH\rightarrow Vb\bar b$ channels, the production depends on the coupling of the
Higgs boson to the weak vector bosons, while the decay is to fermions. In the 
$gg\rightarrow H\rightarrow W^+W^-$ channels, the production is dominated by the
Higgs boson couplings to fermions via the quark loop process, but the decay is 
to bosons. We follow the procedures of the 
LHC Higgs cross section WG~\cite{lhchwg} 
and assuming uniform prior for all coupling's.
Figure~\ref{fig:coupling} shows the constraining of the custodial symmetry 
$\lambda_{WZ}$, the ratio of Higgs coupling to $W$ boson and $Z$ boson, 
by fixing all fermion coupling to SM predictions for the combined Tevatron 
searches for a Higgs mass of 125 geV/c$^2$ and  
two-dimensional constraining of fermion and boson coupling simultaneously by 
fixing to $\lambda_{WZ}$ = 1, respectively. 
The results are consistent with SM predictions. 

\section{Conclusion}
 In conclusion, 
 the final Tevatron results are presented based on full Run II dataset. 
 The Tevatron has achieved SM sensitivity over most of the accessible mass 
 region.
 We observed a broad excess in the mass range of 115 - 140 GeV/c$^2$ 
 relative to background-only 
 hypothesis with a local p-value of 3 sigma, consistent with 
 the discovery of a Higgs boson at 125 GeV/c$^2$. 
 Studies of the couplings at the Tevatron are consistent with SM predictions 
 and are complementary to those performed at LHC. 

\section*{Acknowledgments} 
We would like to thank the organizers of the XLVIIIth Rencontres de Moriond
electroweak session for a wonderful conference with excellent presentations 
and the CDF and D0 Collaborations for the results presented.


\section*{References}

\begin{thebibliography}{99}
\bibitem{higgs} F. Englert and R. Brout, \Journal{\PRL}{13}{321}{1964};
P.W. Higgs, \Journal{\PRL}{13}{508}{1964};
G.S. Guralnik, C.R. Hagen, and T.W.B. Kibble, \Journal{\PRL}{13}{585}{1964}.
\bibitem{tevhiggs} The CDF and D0 Collaborations, Higgs Boson Studies at the 
, arXiv:1303.6346.
\bibitem{ewkfit}The LEP Electroweak Working Group, Status of March 2012, 
http://lepewwg.web.cern.ch/LEPEWWG/.
\bibitem{mtop} The CDF and D0 Collaborations, Combination of the top-quark mass
measurements from the Tevatron collider, arXiv:1207.1069.
\bibitem{mw}The CDF and D0 Collaborations, 2012 Update of the Combination of 
CDF and D0 Results for the Mass of the $W$ Boson, arXiv:1204.0042.
\bibitem{lep}R. Barate {\it et al}, \Journal{\PLB}{565}{61}{2003}.
\bibitem{atlas}G. Aad {\it et al}, \Journal{\PLB}{716}{1}{2012}.
\bibitem{cms}S. Chatrchyan {\it et al}, \Journal{\PLB}{716}{30}{2012}.
\bibitem{tevbb}T. Aaltonen {\it et al}, \Journal{\PRL}{109}{071804}{2012}.
\bibitem{xsec} Tev4LHC higgs Working Group, Standard Model Higgs cross 
sections at hadron colliders, 
http://maltoni.home.cern.ch/maltoni/TeV4LHC/SM.html.
\bibitem{hbr}S. Dittmaier {\it et al},
arXiv:1201.3084. 
\bibitem{VZxsec} J.M. Campbell and R.K. Ellis, 
\Journal{\PRD}{60}{113006}{1999}. 
\bibitem{wwd0}V.M. Abazov {\it et al}, arXiv:1303.0823.
\bibitem{hcp2012}A. Juste, Proceedings of 
Hadron Collider Physics Symposium 2012, 
http://http://www.icepp.s.u-tokyo.ac.jp/hcp2012/
\bibitem{nunubb}T. Aaltonen {\it et al} \Journal{\PRD}{87}{052008}{2013}. 
\bibitem{lhchwg}A. David {\it et al}, arXiv:1209.0040.   

\end{thebibliography}
\end{document}